# Concrete Syntax with Black Box Parsers

Rodin T. A. Aarssen[a,b], Jurgen J. Vinju[a,b], and Tijs van der Storm[a,c]

a Centrum Wiskunde & Informatica; Amsterdam, The Netherlands

b Eindhoven University of Technology; Eindhoven, The Netherlands

c University of Groningen; Groningen, The Netherlands

**Abstract** *Context:* Meta programming consists for a large part of matching, analyzing, and transforming syntax trees. Many meta programming systems process abstract syntax trees, but this requires intimate knowledge of the structure of the data type describing the abstract syntax. As a result, meta programming is error-prone, and meta programs are not resilient to evolution of the structure of such ASTs, requiring invasive, fault-prone change to these programs.

*Inquiry:* Concrete syntax patterns alleviate this problem by allowing the meta programmer to match and create syntax trees using the actual syntax of the object language. Systems supporting concrete syntax patterns, however, require a concrete grammar of the object language in their own formalism. Creating such grammars is a costly and error-prone process, especially for realistic languages such as Java and C++.

*Approach:* In this paper we present Concretely, a technique to extend meta programming systems with pluggable concrete syntax patterns, based on external, black box parsers. We illustrate Concretely in the context of Rascal, an open-source meta programming system and language workbench, and show how to reuse existing parsers for Java, JavaScript, and C++. Furthermore, we propose Tympanic, a DSL to declaratively map external AST structures to Rascal's internal data structures. Tympanic allows implementors of Concretely to solve the impedance mismatch between object-oriented class hierarchies in Java and Rascal's algebraic data types. Both the algebraic data type and AST marshalling code is automatically generated.

*Knowledge:* The conceptual architecture of Concretely and Tympanic supports the reuse of pre-existing, external parsers, and their AST representation in meta programming systems that feature concrete syntax patterns for matching and constructing syntax trees. As such this opens up concrete syntax pattern matching for a host of realistic languages for which writing a grammar from scratch is time consuming and error-prone, but for which industry-strength parsers exist in the wild.

*Grounding:* We evaluate Concretely in terms of source lines of code (SLOC), relative to the size of the AST data type and marshalling code. We show that for real programming languages such as C++ and Java, adding support for concrete syntax patterns takes an effort only in the order of dozens of SLOC. Similarly, we evaluate Tympanic in terms of SLOC, showing an order of magnitude of reduction in SLOC compared to manual implementation of the AST data types and marshalling code.

*Importance:* Meta programming has applications in reverse engineering, reengineering, source code analysis, static analysis, software renovation, domain-specific language engineering, and many others. Processing of syntax trees is central to all of these tasks. Concrete syntax patterns improve the practice of constructing meta programs. The combination of Concretely and Tympanic has the potential to make concrete syntax patterns available with very little effort, thereby improving and promoting the application of meta programming in the general software engineering context.

## 1 Introduction

Meta programming refers to the art of writing programs that analyze, manipulate, or generate source code. Examples of meta programs include compilers, type checkers, obfuscators, interpreters, static analyzers, and many others. Many meta programming systems use Abstract Syntax Trees (ASTs) to represent source code. Meta programming then consists for a large part of visiting, analyzing, and transforming ASTs. Unfortunately, this requires intimate knowledge of the data type or class hierarchy defining the abstract syntax. This problem is especially severe in a reverse engineering context where meta programming is applied to analyze large bodies of code written in "large" languages, such as COBOL, PHP, Java, and C++. The abstract syntax of such languages often consists of more than hundred constructors, divided over multiple syntactic categories.

Concrete syntax patterns [2, 7, 24] alleviate this situation by allowing meta programs to match and create syntax trees using the actual concrete syntax of the object language. As a result, such patterns are less dependent on constructor names or class names, and avoid having to deal explicitly with "empty nestings" as a result of chain rules in the syntax. Unfortunately, systems supporting concrete syntax patterns, such as Asf+Sdf [7], Rascal [18], and Spoofax [15], require a concrete grammar of the object language in their respective formalisms for parsing concrete fragments. Developing such grammars is a costly and error-prone endeavor, especially for real-life programming languages such as Java [17].

In this paper we present Concretely, a simple but effective technique to extend meta programming systems with concrete syntax patterns using black box parsers. Given a mapping from the external parser's syntax tree structures to the meta programming system's internal data structures, Concretely allows meta programmers to plug such external parsers into the pattern matching engine of a meta programming system. As a result, tried and proven off-the-shelf parsers can be reused as is, while gaining the benefits of concrete syntax patterns.

We have implemented Concretely in Rascal, an open-source meta programming language and language workbench [18][1]. We evaluate the technique by connecting external parsers for Java, C++, JSON, and JavaScript to Rascal's pattern matching engine, and assessing the effort required. These results show that, given a mapping from external AST to internal AST, supporting concrete syntax patterns requires very little effort.

Concretely requires marshalling the native ASTs from the external parser to internal Rascal ASTs. Writing such mappings by hand is tedious and error-prone. To alleviate this situation, we introduce Tympanic, a declarative language to define a mapping between a Java class hierarchy and an algebraic data type (ADT) in Rascal. The data type itself is inferred from a Tympanic specification, and all marshalling code is automatically generated.

The contributions of this paper can be summarized as follows:

[1] https://www.rascal-mpl.org (last accessed on 2019-01-30).

- We motivate the benefits of concrete syntax patterns for meta progamming and identify the primary challenges in reusing black box parsers for its implementation (Section 2).
- We present CONCRETELY, a lightweight API in Rascal for supporting concrete syntax using external parsers, and discuss its implementation (Section 3).
- We present TYMPANIC, a declarative mapping language to automatically obtain Rascal ADTs and marshalling code from external AST to internal AST (Section 4).
- We evaluate CONCRETELY using concrete syntax bindings for Java, JavaScript and C++, reusing existing parsers, and qualitatively assess the effort it took to create them (Section 5.1). Futhermore, we show the benefits of using TYMPANIC in terms of code reduction compared to manual implementation of AST marshalling code (Section 5.2).

The paper concludes with a discussion of limitations, directions for further research, and related work (Section 6).

## 2 Concrete Syntax Patterns

**Figure 1** Matching nullary void C++ functions: abstract (left) vs concrete (right).

```
1 void printNullaryFunctions(Program ast) {
2   visit(ast) {
3     case functionDefinition([],
4          declSpecifier([], [],\void()),
5          functionDeclarator([], [], [],
6            name(str n), [], []), [],
7            compoundStatement([], [*_])):
8       println(n);
9 }
```

```
1 void printNullaryFunctions(Program ast) {
2   visit(ast) {
3     case (Decl)`void <Name n>() {<Stm* _>}`:
4       println(n);
5 }
```

### 2.1 Motivation

Analyzing and transforming source code involves making many case distinctions between different kinds of syntax constructs, and decomposing syntax trees in their constituent parts. In the following we take the functional programming perspective, using Rascal as the language to express examples. Note, however, that our observations about AST matching and construction apply also to other functional programming languages or transformation systems with pattern matching, and object-oriented traversal idioms using, e.g., the Visitor pattern [11, 19].

As an example, consider a meta program to print out the names of nullary void functions in C++ code. Figure 1 shows two variants of this meta program in Rascal. The left shows the function `printNullaryFunctions` using idiomatic AST pattern matching. Rascal's built-in traversal operator **visit** traverses the AST and tries to match the pattern against every subtree. If it matches, the variable `n` will be bound to the function's name, and is printed. The right of Figure 1 shows an equivalent function, this time using a concrete syntax pattern. Again, the AST is traversed, but the visit-case now uses the actual surface syntax of C++ within back ticks. The pattern uses typed concrete syntax holes (between angular brackets < and >) to match out the name of the function, and to ignore the statements in the body.

**Table 1** Number of AST types and constructors per language, as they are defined in Rascal.

| Language | #Types | #Constructors |
|---|---|---|
| JavaScript[2] | 3 | 113 |
| PHP[3] | 34 | 171 |
| Java[4] | 7 | 152 |
| C++[5] | 11 | 345 |

Comparing both versions of the function, we can make the following observations. First of all, the abstract version is much larger than the concrete version, and contains more syntactic line noise (parentheses, commas, brackets, etc.). In general, the concrete patterns are much more what-you-see-is-what-you-get (WYSIWYG), which improves readability of the meta program.

Second, the abstract version is highly dependent on names. Using abstract syntax matching the programmer needs to know the exact constructor names of each relevant syntactic construct of the language. In the concrete syntax version only the top-level syntactic category name is needed (i.e., `Decl`). Next to the names of tree constructors, the programmer needs to know the arity of each constructor. In the concrete case this follows directly from the (familiar) surface syntax of the object language.

Depending on the size of the language, navigating such AST data types can be a daunting task. For reference, Table 1 shows the number of AST types (syntactic categories) and the total number of tree constructors for a number of mainstream languages, as they are defined in Rascal. As can be seen, such AST data types can be quite intimidating.

Third, abstract syntax structures do not support abstracting over implicit injections (or "chain rules"). For instance, to match on the **void** type, the programmer needs to explicitly match the `declSpecifier` construct in the second argument of `functionDefinition`. In the concrete syntax version this is taken care of by the parser.

Finally, the abstract patterns require the absence of information to be explicitly accounted for, either using unit values (e.g., `[]`), or dummy placeholders (e.g., `_`). In the concrete patterns absence of information is often represented by simply omitting elements. For instance, to match only on nullary functions, the pattern simply leaves out any formal parameters.

### 2.2 Implementing Concrete Syntax Patterns

Meta programming systems that support concrete syntax patterns can be divided in two main categories. The first category consists of languages that support concrete syntax over a fixed object language. Examples of this category are Template Haskell [23] and MetaOCaml [8]. We consider these languages out of scope since they do not support meta programming for the purpose of reverse engineering or reengineering [3].

The other category consists of transformation systems which come with their own grammar formalism such as Asf+Sdf [7], TXL [9], Spoofax [15], and Rascal. In all these systems the use of concrete syntax patterns presupposes a correct, non-ambiguous grammar of the object language, defined in the particular system's grammar formalism. Unfortunately, writing correct, non-ambiguous grammars for real programming language is very hard [17].

It would therefore be valuable if existing, off-the-shelf, parsing infrastructure could be used. Many industrial strength parsers for mainstream languages exist in the form of compilers (e.g., Clang, Javac), IDE frameworks (e.g., Eclipse JDT, CDT, Roslyn C#), static analysis toolkits (e.g., SonarQube), or actual parsers defined by third parties (e.g., grammars defined by ANTLR [22]). However, reusing such parsers for concrete syntax patterns presents two main challenges. We discuss each one of them in turn.

**AST marshalling and impedance mismatch** Parsers turn source code into trees. However, the resulting trees of an external parser cannot be reused as is in a meta programming system. First of all, the implementation of the AST or concrete syntax tree (CST) most likely will not match the internal data structures used to represent syntax trees in the meta programming system. For instance, a parser of the Java language might build a tree defined by an object-oriented class hierarchy in Java, whereas, for instance, Rascal uses generic, immutable data structures to represent trees.

Furthermore, when pattern matching syntax trees, we would like to abstract from accidental parser dependent idiosyncrasies in the tree representation. For instance, many LALR parser generators (such as Yacc [14]) do not support the EBNF regular operators ("*", "+", "?") to represent sequences or optionality. As a result, sequencing and optionality is encoded using additional non-terminals which might pollute the trees produced by such parsers, leading, for instance, to a sequence of statements to be represented as a heavily unbalanced tree. Conversely, parsing systems that do not support left recursive productions (e.g., earlier versions of ANTLR [21]) might produce convoluted expression trees which are hard to match.

---

[2] https://github.com/cwi-swat/js-air/blob/master/src/lang/javascript/m3/AST.rsc (last accessed on 2019-01-30).

[3] https://github.com/cwi-swat/php-analysis/blob/master/src/lang/php/ast/AbstractSyntax.rsc (last accessed on 2019-01-30).

[4] https://github.com/usethesource/rascal/blob/master/src/org/rascalmpl/library/lang/java/m3/AST.rsc (last accessed on 2019-01-30) and https://github.com/usethesource/rascal/blob/master/src/org/rascalmpl/library/lang/java/m3/TypeSymbol.rsc (last accessed on 2019-01-30).

[5] https://github.com/cwi-swat/clair/blob/master/src/lang/cpp/AST.rsc (last accessed on 2019-01-30) and https://github.com/cwi-swat/clair/blob/master/src/lang/cpp/TypeSymbol.rsc (last accessed on 2019-01-30).

**Limited parsing capabilities** Parsers in the wild often offer only a small set of start nonterminals, at the level of whole programs or compilation units. Unfortunately, this severely limits the applicability of concrete syntax patterns, where patterns are often applied at the level of arbitrary subtrees, which requires parsing over arbitrary nonterminals. Furthermore, concrete syntax patterns might contain holes to match out or insert subtrees in a pattern. External parsers written for a different purpose will not support the syntax of holes. Addressing both these problems requires invasive modifications to the parser. Sometimes this might not even be possible, because the source code of the parser is unavailable. If it is possible, however, it can be a quite daunting and error-prone task. Most parsers cannot be modularly extended, so existing code must be modified. When a parser is generated from a grammar formalism, it might be slightly easier, but would likely still lead to ambiguities or shift/reduce conflicts, or degradation of performance due to backtracking.

In the next section we assume that the AST marshalling and impedance mismatch problem is addressed by the user of Concretely explicitly. That is, we assume that there is a mapping from the external parser's syntax tree to, in this case, a Rascal AST structure conforming to a Rascal-defined Algebraic Data Type (ADT). In Section 4 we introduce the Tympanic language to obtain such mappings automatically from a declarative specification. The second problem is addressed by Concretely, which we discuss first.

## 3 Concretely in Rascal

### 3.1 Introduction

Concretely provides an interface to Rascal meta programmers to connect external parsers producing ASTs conforming to Rascal data types, leveraging the concrete pattern matching engine on those ASTs. In this section we introduce the API using the example language of JSON.

The abstract syntax and Concretely interface for JSON in Rascal are shown in Figure 2. The abstract syntax is defined by the algebraic data type `JSON`. It defines constructors for booleans, strings, numbers, null, arrays, and objects. An object contains a list of properties. A `Prop` property has an identifier (`Id`) as its name, and a `JSON` value. Identifiers of type `Id` simply wrap a string[6].

[6] Note that we could have defined an object's properties simply as a `map[str, JSON]`, using Rascal's built-in map data structure, but at present the pattern matching engine does not support maps, so we fall back to an explicit list of properties. We revisit this problem in Section 6.

**Figure 2** A Rascal CONCRETELY module for JSON.

```
module JSONConcretely

data JSON
  = boolean(bool b) | number(real n) | string(str s) | array(list[JSON] elts)
  | null() | object(list[Prop] props);

data Prop = prop(Id name, JSON val);
data Id = id(str name);

@concreteSyntax{JSON}
java JSON parse(str json);

@concreteSyntax{Prop}
Prop parseProp(str prop) = parse("{<prop>}").props[0];

@concreteHole{JSON}
str jsonHole(int id) = "{_hole: <id>}";

@concreteHole{Prop}
str propHole(int id) = "_hole: <id>";
```

The remaining part makes up the core of the CONCRETELY API. First, for all (abstract) nonterminals that should be exposed for concrete pattern matching and construction, a parse function needs to be defined, annotated using the tag `@concreteSyntax{T}`, where $T$ represents the nonterminal to be parsed. In this case, when the Rascal interpreter encounters a concrete syntax fragment of type `JSON` (e.g., `(JSON)`...``), the parse function is called to turn the pattern into an abstract pattern conforming to the type `JSON`. Note that the `parse` function has the **java** modifier, which indicates that this function uses Rascal's foreign-function interface (FFI) to call an external black box JSON parser and converter to Rascal's representation of ASTs.

The parse function may contain more logic to interface with the external parser if there are restrictions on the start nonterminals of the parser. For instance, if the external parser can only parse object literals, the parse function could still be defined as follows:

```
@concreteSyntax{JSON}
JSON parseWithContext(str json) = parse("{dummy: <json>}").props[0].val;
```

In this case, the function `parseWithContext` first wraps the arbitrary JSON string in a dummy object literal[7], parses the object literal using the external parser, and then returns the property value corresponding to the input. This same technique is used in the parser for `Prop` fragments, which are wrapped in curly parentheses to enable parsing.

The second part of the CONCRETELY API is the construction of placeholder values for pattern holes. This is realized by the functions `jsonHole` and `propHole`, which create identifiable, dummy JSON strings to represent meta variables in a pattern, without having to modify the external parser. Functions with this purpose are annotated with @concreteHole{*T*}, where *T* represents the nonterminal type of the hole. In this case, a JSON hole is represented by an object literal of the form {_hole:*id*}, where *id* represents a unique number assigned by CONCRETELY. This unique number is used by CONCRETELY to replace these dummy placeholders in the result of the external parser with actual holes before pattern matching or instantiation (cf. Section 3.3).

[7] The angular brackets < and > are used for string interpolation here.

### 3.2 Using CONCRETELY

Given a module such as the one shown in Figure 2, it is now possible to use concrete syntax for matching and construction of JSON terms. Below we show a number of examples using Rascal's command-line interface.

For instance, the following snippet shows how to construct a JSON number term[8]:

```
> age = (JSON)`29`;
>> number(29.0)
```

So the user enters actual JSON syntax within the back ticks, and the result is an AST over the JSON type.

Holes in a pattern can be used to compose more complex terms. The following interaction shows the construction of an object literal interpolating the previously created age value as one of its properties:

```
> p = (JSON)`{name:"Rodin",age:<JSON age>}`;
>> object([prop(id("name"), string("Rodin")), prop(id("age"), number(29.0))])
```

Concrete patterns can also be used to match ASTs. One way to do pattern matching in Rascal is by using the infix := operator. The following snippet uses Rascal's list matching to check for the presence of a name property in a JSON object:

```
> (JSON)`{<Prop* _>, name: <JSON _>, <Prop* _>}` := p;
>> true
```

Effectively, the pattern searches for the first name property in the list of properties in the object, ignoring properties before and/or after. Internally, the concrete pattern corresponds to the pattern object([_*, prop(id("name"), _), _*]).

### 3.3 Implementation

This section describes the implementation of CONCRETELY in the Rascal interpreter. First, we note that we did not change the grammar of Rascal itself; instead, we reuse Rascal's existing notation for concrete syntax fragments, which originally was only used for languages with a concrete grammar.

This notation consists of two main parts: the type of the concrete syntax fragment (the nonterminal), and the concrete syntax itself. As an example, consider the following concrete syntax fragment: (Expr)`1+2`. Here, Expr represents the type, and 1+2 is the actual object language source fragment.

[8] The > character indicates the prompt, and lines starting with >> display the result of evaluation.

| | |
|---|---|
| Concrete pattern | (JSON)`{<Prop p>}` |
| CONCRETELY | ⇓① |
| Textual encoding | "{_hole: 0}" |
| External parser | ⇓② |
| Foreign AST | △ |
| AST marshaller | ⇓③ |
| Rascal AST | object([prop(id("_hole"), number(0.0))])) |
| CONCRETELY | ⇓④ |
| Abstract pattern | object([Prop p]) |

**Figure 3** Steps in the CONCRETELY pipeline.

As seen in the examples, the actual concrete fragment may contain holes <T x> as placeholders for subtrees. These holes are used to match and bind subtrees of a syntax tree or insert trees into the pattern. Holes are always qualified with a type to ensure syntax safety [4].

An overview of the CONCRETELY process is shown in Figure 3. The first step consists of the interpreter recognizing a concrete pattern and confirming that there is an ADT with the right name in scope in the interpreter environment (e.g., the JSON data type of the module JSONConcretely must be visible). Using the functions annotated with concreteSyntax and concreteHole, the concrete pattern is then converted to an encoded representation in terms of the ADT (e.g., JSON).

Steps ①, ②, and ③ are encapsulated by the concreteSyntax-annotated parse function: the textual encoding is parsed by the external parser, the output of which is converted to a Rascal AST.

CONCRETELY then parses the hole placeholders with the applicable concreteSyntax-annotated parse function to obtain their Rascal equivalent, and replaces occurrences in the obtained AST with the appropriate hole subtree to turn the resulting AST into a proper abstract pattern, after which Rascal's ordinary pattern evaluation logic takes over. This conforms to Step ④.

Thus, CONCRETELY reuses the annotated parsers to parse and retrieve the encoded hole fragments as well. So whenever a function is annotated with concreteHole for some non-terminal $T$, this requires a parser for $T$ as well. For instance, in Figure 2, a Prop hole is encoded as "_hole:$n$", but such a string cannot be parsed as is. Therefore, the function parseProp first wraps such fragments in an object literal and then retrieves the nested Prop value.

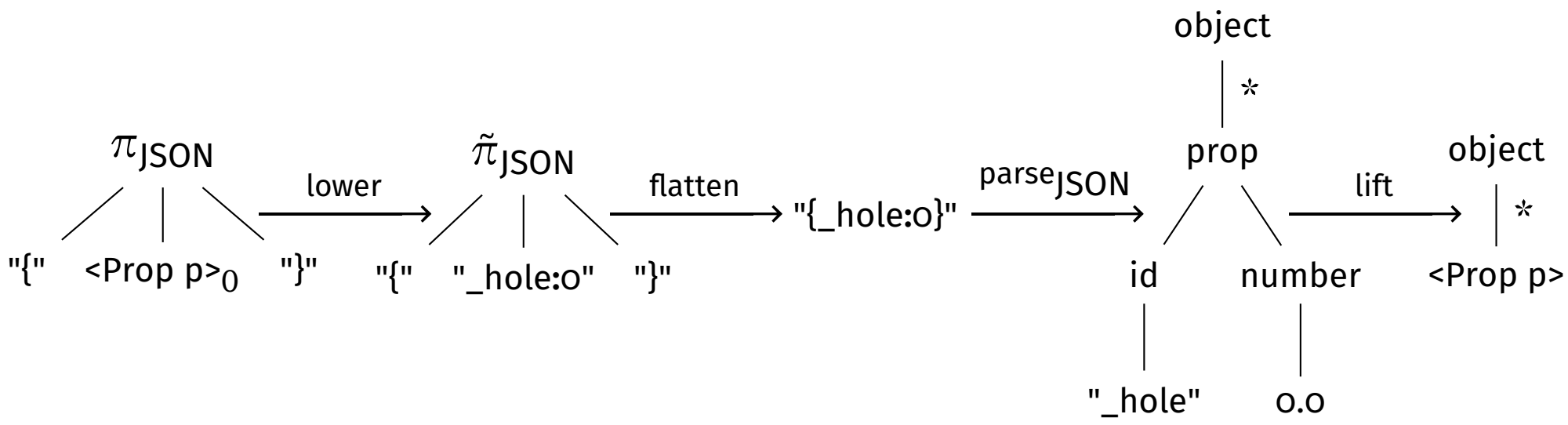

**Figure 4** Intermediate representations when processing (JSON)`{<Prop p>}`.

## 3.4 Formal Description

This section formally describes the individual steps and intermediate representations in the Concretely pipeline. Let $\mathsf{parse}_T$ and $\mathsf{hole}_T$ be the parsing and hole encoding functions for type T (cf. Section 3.3). To encode hole meta variables in the object language, we define a mapping lower. First, we index all holes <T x> in a term with a unique number $i$. Then, each hole $\langle T\ x\rangle_i$ is mapped to $\mathsf{hole}_T(i)$. The lowered hole is subsequently parsed with $\mathsf{parse}_T$ to obtain its AST representation. The inverse mapping lift then maps the obtained AST fragment back to the original hole <T x>. The mappings lower and lift are defined as follows.

$$\mathsf{lower} = \bigcup_{\langle T\ x\rangle_i \in \mathsf{pattern}} \langle T\ x\rangle_i \mapsto \mathsf{hole}_T(i) \qquad \mathsf{lift} = \bigcup_{\langle T\ x\rangle_i \in \mathsf{pattern}} \mathsf{parse}_T(\mathsf{hole}_T(i)) \mapsto \langle T\ x\rangle$$

As an example, consider the pattern of Figure 3. This term describes a JSON object, containing a hole for a single Prop (cf. Figure 2). Given the parse and hole functions from Figure 2, the mappings lower and lift are as follows:

$$\mathsf{lower} = \{\langle \mathsf{Prop}\ p\rangle_0 \mapsto \text{"_hole:0"}\}$$

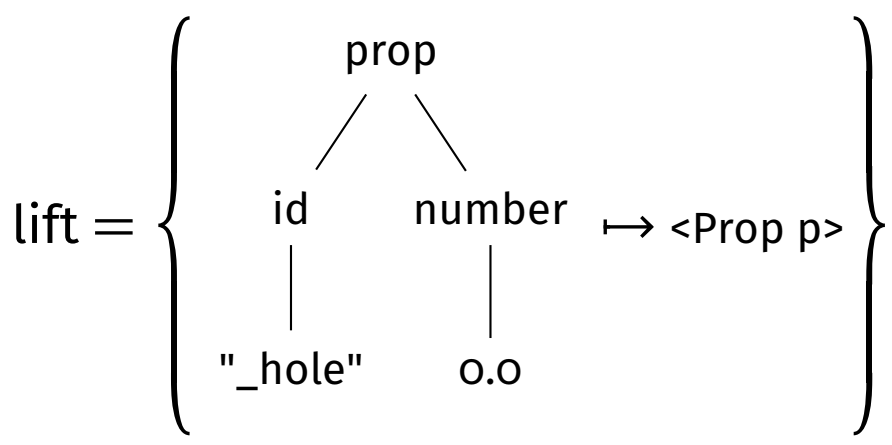

Figure 4 illustrates intermediate representations and steps in detail. Internally, the pattern is translated into a tree over $\pi_{\mathsf{JSON}}$, the type of concrete patterns over JSON. This tree has three leaves for its constituent parts, namely the string values "{" and "}", with a subtree representing the hole $\langle \mathsf{Prop}\ p\rangle_0$ in the middle. The lower mapping transforms this tree into a tree over $\tilde{\pi}_{\mathsf{JSON}}$, the type of concrete patterns over JSON without holes. This tree again has three leaves, namely the string values "{", "_hole:0", and "}". The intermediate tree is then flattened to a string by concatenating the leaves. The resulting JSON fragment "{_hole:0}" is input to $\mathsf{parse}_{\mathsf{JSON}}$, resulting in a JSON AST with encoded holes. Finally, the lift mapping turns this tree back into a proper pattern, this time over the abstract syntax of JSON.

## 4 Tympanic: Mapping AST Class Hierarchies to Algebraic Data Types

### 4.1 Introduction

External parsers for programming languages come with their own AST representation for the object language, for instance in the form of an object-oriented class hierarchy in Java. Meta programming systems, like Rascal, on the other hand, might employ incompatible paradigms (e.g., class hierarchies vs algebraic data types) to describe abstract syntax. Using an external parser's AST in a specific meta programming system, requires marshalling code to transform an external AST into an internal one, thereby solving a potentially deep impedance mismatch [13]. This task can be daunting for two reasons: first, it requires the definition of an abstract grammar on the meta side; second, it involves writing a transformation that traverses the external AST in order to build up an internal AST.

In this section we introduce Tympanic[9], a mapping DSL that improves this situation for the case of Java class hierarchies and Rascal's algebraic data types. Tympanic supports defining a declarative mapping from Java classes and Rascal data types, from which the abstract grammar and marshalling code is automatically generated.

### 4.2 Syntax

Figure 5 shows the (slightly simpified) syntax definition of Tympanic[10]. A Tympanic mapping starts with declaring a list of imported Java packages to provide access to the native ASTs of an external parser. The `export` clause specifies the qualified name of the Rascal module that will be generated with the ADT.

The `types` section provides a mapping from Java types (classes and interfaces) to Rascal ADTs. Tympanic assumes that a subset of all Java types will map to Rascal ADTs; multiple Java types can map to the same ADT. This section is used, for instance, to map a Java interface `Expr` to a corresponding Rascal ADT `Expr`, by declaring `Expr ⇒ Expr`.

The final section defines how concrete objects in a Java AST will be mapped to constructors of the ADTs introduced in the `types` section. Every constructor mapping starts with the declaration of a concrete AST class, followed by one or more patterns on a Java object, binding and restricting certain Java fields or getters (both captured by the `JavaField` nonterminal). In each pattern, the Java fields are positionally mapped to the arguments of the constructor application after the colon. The types of the arguments are inferred by analyzing the type hierarchy of Java and the mapping specification of the `types` section.

Since in many cases the fields/getters from a Java class do not match directly to constructor arguments, fields can be guarded, by restricting their value (e.g., `hasBody == true`, or `getOp == Operators.PLUS`), declaring them optional (e.g., `getElse?`), or restricting to some

[9] The name is inspired by the tympanic cavity in the middle ear, which solves the impedance mismatch between air and liquid.

[10] We use Rascal's built-in notation for context-free grammars, which mostly corresponds to standard EBNF. A symbol $\{S\ \ s\}*$ indicates a repetition of $S$, with $s$ as a separator.

**Figure 5** Syntax definition of Tympanic (simplified).

```
1  start syntax ASTMapping
2    = "mapping" Id Import* "export" {Id "::"}+ "types" Datatype* "constructors" Mapping*;
3
4  syntax Import = "import" {Id "."}+;
5
6  syntax Datatype = Id "=\>" Id;
7
8  syntax Mapping = Id Match+
9
10 syntax Match = "-" {JavaField ","}* ":" Constructor;
11
12 syntax JavaField = Field | "%" Field;
13
14 syntax Field
15   = Id | Id "==" JavaValue | Id "!=" JavaValue | Id "?"
16   | "(" Id ")" Id | "(" Id "[" "]" ")" Id;
17
18 syntax Constructor = Id "(" {Arg ","}* ")";
19
20 syntax Arg = Id | Id Id "=" RascalValue;
21
22 syntax JavaValue = "null" | "true" | "false" | Int | {Id "."}+;
23
24 syntax RascalValue = "true" | "false" | Int | Id "(" {RascalValue ","}* ")";
```

subtype using Java's cast notation (e.g., `(Integer)getValue`). Finally, if a field/getter is used solely for dispatching but should not end up in the list of constructor arguments, it can be skipped using `%`. Optional fields indicated by `?` map to `Maybe` values in Rascal, since Rascal does not have null-values.

### 4.3 An Example Tympanic Specification

An example Java AST class hierarchy and corresponding Tympanic mapping specification is shown in Figure 6; the resulting Rascal ADT is shown in Figure 7. The class hierarchy on the right of Figure 6 uses an enum to encode binary operators of the `Binary` class. In the mapping each variant is mapped to individual constructors, `add`, `mul`, `sub`, and `div`. Since the operator retrieved from `getOp` is not used in the constructor, it is skipped using `%`.

The `Cond` class implements a conditional expression, but in this case it is known that the else-branch is optional. Both variants are distinguished using comparison guards on `getElse`, mapping to two different AST constructors.

The `Block` class wraps a sequence of expressions, represented as a Java array. Arrays and arguments (indirectly) implementing `java.util.Iterable` are mapped to Rascal lists, as illustrated on line 6 of Figure 7.

Finally, the `Lit` class wraps arbitrary Java objects as literal expressions. The Tympanic mapping specifies three supported types, which are dispatched using the cast notation.

**Figure 6** Example AST class hierarchy (left) and TYMPANIC mapping specification (right).

```
package expressions;

enum Op {
  PLUS,
  TIMES,
  MINUS,
  SLASH
}

interface Expr { }

class Binary implements Expr {
  Expr getLhs() {...}
  Expr getRhs() {...}
  Op getOp() {...}
}

class Cond implements Expr {
  Expr getCond() {...}
  Expr getThen() {...}
  Expr getElse() {...}
}

class Block implements Expr {
  Expr[] getBody() {...}
}

class Lit implements Expr {
  Object getValue() {...}
}
```

```
mapping ExprAst

import expressions

export expr::Expr

types Expr ⇒ Expr

constructors

Binary
  - %getOp == Op.PLUS, getLhs, getRhs: add(lhs, rhs)
  - %getOp == Op.TIMES, getLhs, getRhs: mul(lhs, rhs)
  - %getOp == Op.MINUS, getLhs, getRhs: sub(lhs, rhs)
  - %getOp == Op.SLASH, getLhs, getRhs: div(lhs, rhs)

Cond
  - getCond, getThen, %getElse == null: ifThen(cond, then)
  - getCond, getThen, getElse != null: ifThenElse(cond, then,
        ↪ els)

Block
  - getBody: block(body)

Lit
  - (Integer)getValue: integer(intVal)
  - (Boolean)getValue: boolean(boolVal)
  - (String)getValue: string(strVal)
```

**Figure 7** Generated Rascal module from the TYMPANIC mapping of Figure 6.

```
module expr::Expr

data Expr
  = add(Expr lhs, Expr rhs) | mul(Expr lhs, Expr rhs) | sub(Expr lhs, Expr rhs) | div(Expr lhs, Expr rhs)
  | ifThen(Expr cond, Expr then) | ifThenElse(Expr cond, Expr then, Expr els)
  | block(list[Expr] body)
  | integer(int intVal) | boolean(bool boolVal) | string(str strVal);
```

In this case, the marshalling code generated by TYMPANIC will convert the Java primitive types to corresponding Rascal values of type **int**, **bool**, and **str**, respectively.

As an alternative to dispatching each constant of an enum to a unique constructor, enums can be mapped to inferred Rascal ADTs using the `Type arg = ...` notation for a Rascal argument. For instance, the following constructor mapping would implicitly introduce the `Op` data type:

```
Binary
  - getOp == Operator.PLUS, getLhs, getRhs: binary(Op op = plus(), lhs, rhs)
```

In this case, TYMPANIC will generate the following ADTs:

```
data Expr = binary(Op op, Expression lhs, Expression rhs);
data Op = plus();
```

### 4.4 Implementation

Tympanic has been implemented in Rascal, and consists of two components, generating the Rascal ADT and the marshalling Java code, respectively. Both compilation steps rely on type analysis of the Java class hierarchy. Tympanic uses the JDT binding to the $M^3$ meta model [6] to inspect the Java type inheritance relation in order to map constructors to their nonterminal types, and to obtain the types of fields and getters.

In the first compilation step, the Tympanic specification is compiled to the corresponding abstract Rascal grammar, as illustrated in Section 4.3. The `types` section is used to obtain the ADT type corresponding to the constructor in a constructor mapping. To obtain the type of a constructor argument, the return type of the field or getter is found through the $M^3$ model.

The second compilation step yields a marshaller, which performs the conversion from the external AST to a value corresponding to the ADT generated in the first step. The marshaller is essentially a visitor over the external AST, with entry points for all nonterminals in the specification. For each of the nonterminals, the compiler detects which mapping rules apply to a subtype of the external type, and generates an **instanceof** switch over all possibilities. All relevant mapping constraints (cf. Section 4.2) are then evaluated to ensure that the rule applies. Finally, all external arguments are evaluated and visited, so that the Rascal constructor value can be built and returned.

In total the Tympanic implementation consists of ±500 source lines of Rascal code[11]. Section 5.2 discusses the benefits in terms of effort when using Tympanic compared to manual implementation of AST marshalling.

## 5 Evaluation

### 5.1 Benefits of Concretely

In Section 3.1, we described the necessary components to implement concrete syntax pattern matching, and illustrated the technique with the JSON language. In this section we evaluate Concretely in the context of three mainstream programming languages, C++, Java, and JavaScript.

Table 2 shows an overview of each of these Concretely instantiations, estimating effort in terms of source lines of code (SLOC). The C++ binding is based on the Eclipse CDT parser. C++ is a large language, requiring 7232 SLOC of mapping code;

[11] https://github.com/cwi-swat/tympanic/tree/f0e3b454289204a55ba637b4bea196fa0490411d (last accessed on 2019-01-30).

**Table 2** Evaluation of Concretely in terms of SLOC.

| Language | Parser | ADT | Mapping | Binding |
|---|---|---|---|---|
| C++[12] | Eclipse CDT | 387 | 7232 | 55 |
| Java[13] | Eclipse JDT | 183 | 1806 | 49 |
| JavaScript[14] | Nashorn | 130 | 429 | 22 |

**Figure 8** An excerpt from Rascal's abstract C++ grammar.

```
1  data Declaration
2    = translationUnit(list[Declaration] declarations)
3    | functionDefinition(Expression returnSpec, Declarator declarator, Statement body)
4    | ...
5    ;
6
7  data Statement
8    = compoundStatement(list[Attribute] attributes, list[Statement] statements)
9    | ...
10   ;
```

in this case, this has been partially generated from the C++ ADT specification. The actual Concretely code is limited to 55 SLOC, demonstrating that, given a mapping from foreign AST to Rascal AST, setting up Concretely requires very little effort.

In Rascal, the abstract grammar for C++ consists of more than 300 constructors, over 11 data types. A small excerpt of the abstract syntax data type is shown in Figure 8, showing snippets of the Declaration and Statement types. In contrast to JSON, the C++ syntax has many different syntactic categories, distinguishing, for instance, statements, expressions, and declarations. The external C++ parser, however, only accepts whole C++ programs as input. Because of this, the parsing function not only has to call the external parser, but has to provide some context code for the concrete syntax fragment, such that the parser is provided with a full program.

This is illustrated using the statement parser shown in Figure 9. The concrete syntax fragment of a statement is inserted into a function definition. The resulting complete program is then given to the external parser as input. The (image of the) context has to be stripped off again from the resulting translationUnit AST (cf. Figure 8), which produces the image of the concrete syntax fragment.

The Java binding follows a similar pattern, this time reusing the Eclipse JDT presentation parser as backend. The mapping to Rascal AST is smaller, because there are fewer constructors and nonterminals in Java's abstract syntax. The Concretely code is of comparable size, since for both languages, we have implemented bindings for 5 nonterminals.

[12] https://github.com/cwi-swat/clair (last accessed on 2019-01-30).
[13] https://github.com/usethesource/rascal-eclipse/tree/master/rascal-eclipse/src/org/rascalmpl/eclipse/library/lang/java/jdt (last accessed on 2019-01-30).
[14] https://github.com/cwi-swat/js-air (last accessed on 2019-01-30).

**Figure 9** Concretely code for C++ (excerpt).

```
1 @concreteSyntax{Statement}
2 Statement parseStatement(str stmt) {
3   str prog = "void dummy() { <stmt> }";
4   Declaration decl = parseFromString(prog);
5   return decl.declarations[0].body.statements[0];
6 }
```

**Table 3** Comparing SLOC of Tympanic vs manual implementation of C++ mapping. In the Tympanic mapping, the ADT and Marshaller code is generated from the Tympanic specification.

| Mapping | Tympanic | ADT | Marshaller |
|---|---|---|---|
| Manual | – | 387 | 7232 |
| Tympanic[15] | 411 | 308 | 4433 |

JavaScript is a much smaller language than Java and C++ in terms of syntax. The SLOC numbers are therefore much lower for the ADT and mapping code. The Concretely binding code is roughly of the same order of magnitude as the others.

### 5.2 Tympanic vs Manual Implementation

We have defined a mapping for the Eclipse CDT C++ parser using Tympanic; the results in terms of lines of code are shown in Table 3. The grammar and marshaller that are compiled from this Tympanic specification are equally expressive as the manual implementation discussed in Section 5.1, although there are some differences.

First, next to the grammar and marshaller code that make up the figures in Table 2, the manual implementation contains additional code for name and type resolution. Since this functionality is not sufficiently generalizable, this is not included in Tympanic. Second, the manual implementation does not inline several commonly occurring fields (such as `static`, `const`, and `signed`) into every constructor, but are bundled in a list of modifiers, to improve readability. Finally, instead of using `Maybe` values (cf. Section 4.2) to encode optionals, the manual implementation uses constructor overloading instead.

Table 3 shows an overview of both the manual and the Tympanic-generated mappings in terms of SLOC. Note that for the Tympanic specification, the ADT and marshalling code are generated from the Tympanic specification, whereas for the manual implementation, the ADT was written by hand, and the marshalling code was manually written or generated using a one-off approach.

[15] https://github.com/cwi-swat/tympanic/blob/f0e3b454289204a55ba637b4bea196fa0490411d/src/tympanic/cdt.tymp (last accessed on 2019-01-30).

## 6 Discussion and Related Work

The results shown in Table 2 show that, given an external black box parser and mapping from foreign AST to Rascal AST, making a language eligible for concrete syntax matching and construction requires little effort. Nevertheless, there are some limitations and further directions for Concretely.

### 6.1 Limitations and Further Directions

**Hole Capture** First of all, the user-defined encoding of holes should be unique. If there are ordinary terms that look like holes, the pattern matching engine might incorrectly turn a term into a pattern. For instance, consider the case that the user expresses the following JSON term.

```
(JSON)`{_hole: 0}`
```

Concretely will (correctly) not identify this as a hole, since the lift mapping is empty (cf. Section 3.4). However, when there are metavariables in the pattern, this can lead to confusing ambiguities. For instance, the following expression will be (accidentally) turned into a non-linear match:

```
(JSON)`[<JSON x>, {_hole: 0}]` // ⇒ array([<JSON x>, <JSON x>])
```

Currently, it is up to the users of Concretely to ensure that such accidental captures of holes are unlikely to happen, for instance, by using sufficiently atypical names and patterns. A possible solution for improving this is to use the source locations (e.g., filename and character offset) of the holes in the meta program as part of the unique representation. The likelihood of such values to be accidentally occurring in object terms is very small.

**Built-in Datatypes** The current implementation of Concretely only supports matching and construction of algebraic data type values and lists. Rascal, however, supports additional built-in data types, such as primitive types for integers, strings, booleans etc., and a range of collection types, such as sets, $n$-ary relations, and maps. It would be interesting to see if Concretely can be extended to support these types in concrete syntax patterns as well.

For instance, consider the representation of properties in JSON object literals of Figure 2. Currently this is defined as a list of `Prop` values, which supports matching against property ASTs. Alternatively, the `object` constructor could also have been defined as follows:

```
data JSON
  = ... | object(map[str, JSON] props);
```

Here, an object literal simply contains a map from string values to JSON values. The concrete syntax pattern syntax would then have to change to support patterns like the following:

```
(JSON)`{<str name>: <JSON _>} := someJson`
```

The pattern now contains a primitive hole of type **str**, to match out the name of a property. Furthermore, the pattern matching engine would also have to be extended to match a sequence of syntactic elements to the actual **map** structure produced by the parser.

### 6.2 Related Work

The relation between concrete syntax (as defined by grammars or parsers) was first explored in the Popart language [26] and the Syntax Definition Formalism (SDF) [12]. The Popart language processing system offers an automatic conversion from a concrete Yacc [14] specification to an abstract BNF-style grammar. The SDF grammar formalism explored the relation from an algebraic specification perspective. The definition of SDF includes an automatic mapping from concrete grammars and algebraic signatures on the one hand, and concrete syntax trees and algebraic terms, on the other hand. In the context of CONCRETELY, this mapping needs to be provided explicitly, for each black box parser that is used.

SDF formed one of the two pillars of the ASF+SDF system [7, 16]. ASF+SDF is unique in that it *only* supports concrete syntax patterns, with an extremely minimal meta language, consisting of conditional rewrite rules. These rewrite rules operate on concrete syntax trees (parse trees) directly. The syntax of holes would be explicitly defined by the programmer (in the SDF's variables section), and the grammar would be merged with ASF's rewrite rule skeleton syntax.

A mapping similar to the one defined by SDF is used in the Stratego transformation system [25], where it is called implode. Stratego has been extended later to support concrete syntax patterns [24]. To use concrete syntax in Stratego the object language grammar is merged with the meta language, with additional rules for meta variables (holes), similar to ASF+SDF's approach. A meta program is then parsed with the extended grammar in one go, and the patterns are imploded to abstract patterns, like in CONCRETELY.

Stratego and CONCRETELY have similarities in approach. For instance, the grammar attributes ToTerm and FromTerm are similar to CONCRETELY's concreteHole and concreteSyntax annotations. In CONCRETELY, however, they apply to functions defined by the user, whereas in Stratego they represent embedding rules from meta language to object language and vice versa.

Integrating concrete syntax as part of a functional programming language was explored earlier in the work of Aasa, Petersson, and Dynek [1, 2]. Programmers can define *conctypes*, algebraic data types defined with concrete grammar rules. Again, however, this requires a built-in grammar formalism and a parser that is able to parse the combined syntax of the meta language and object language terms. Rascal's standard concrete syntax feature uses a two stage approach: first the patterns are parsed as strings with holes (similar to $\pi_{\mathsf{JSON}}$ in Figure 4), which are then parsed using an augmented object language grammar with (generic) placeholders for the holes. CONCRETELY has removed the requirement of the presence of such a concrete grammar.

TXL [9, 10] is a source code transformation system which employs rewrite rules using concrete syntax. Its implementation uses a variant of LL(1) parsing, with ordered alternatives. As a result, new alternatives to a production rule can be added incrementally to extend a base grammar. Source code is then sequentially parsed by all overrides, yielding ASTs for matching and rewriting. Unlike in other approaches, there is no explicit definition of the abstract syntax in the form of algebraic data type or signature.

Object-relational mapping (ORM) is a technique to make use of relational databases in object-oriented programming languages [13]. A programmer defines a mapping between database tables and object structures. This mapping then provides a programmatic interface to the underlying database. ORM is notoriously known for the problem of impedance mismatch, resulting from the differences between the underlying paradigms of object-oriented languages and relational databases.

Lämmel and Meijer [20] give an overview of mappings, mapping concepts, and various instances of impedance mismatches among major data modeling and processing paradigms. Tympanic solves the impedance mismatch problem between class hierarchies and algebraic data types, both representing ASTs for a language.

Migrating from an old API to a new API is a challenging task, even when the APIs are very similar. Bartolomei, Czarnecki, Lämmel, and Van der Storm [5] describe several approaches to migrate between the JDOM and DOM APIs for XML, of which using grammar-based API protocols is the most satisfactory solution. Writing a Tympanic specification could similarly be seen as writing annotations on a foreign API.

## 7 Conclusion

Concrete syntax patterns provide a convenient way to match, decompose, and construct syntax trees. However, transformation systems with concrete object syntax support often require a hand-crafted grammar in the system's own formalism, which can be too costly to construct. In this paper, we have presented Concretely, a lightweight approach to allow the use of concrete syntax patterns by reusing external, black box parsers. The programmer writes concrete syntax patterns, but pattern matches and constructs abstract syntax trees under the hood.

Concretely requires meta programmers to define custom parse functions for each nonterminal of the abstract syntax of a language, calling out to the external parser and converting its native AST to the desired AST structure in the meta programming system. Additionally, the programmer provides functions to encode placeholders (meta variables) of a pattern into a form that can be understood by the external parser, and is sufficiently unique for Concretely to translate back to a proper abstract pattern.

We have implemented Concretely in Rascal, a meta programming system and language for source code analysis and manipulation [18]. Using this prototype implementation we were able to define concrete syntax patterns for industrial programming languages (Java, JavaScript, and C++) using only a few dozen lines of code. Furthermore, to alleviate the burden of writing the mapping between an external parser's

AST structures and Rascal's internal representation, we proposed Tympanic, a DSL for declaratively mapping Java class hierarchies to algebraic data types. Both the algebraic data type and marshalling code is generated automatically, realizing an order of magnitude reduction in lines of code compared to manual implementation.

Directions for further research include supporting a mixture of primitive data type matching/construction in concrete patterns, making Concretely more robust against accidental capturing of placeholders, and parsing patterns at compile-time for early detection of syntax errors.

**Acknowledgements** The first author was supported by NWO grant BISO.15.04: "Model Extraction for Re-engIneering Traditional Software (MERITS)", in collaboration with Philips Healthcare, Best, The Netherlands.

## About the authors

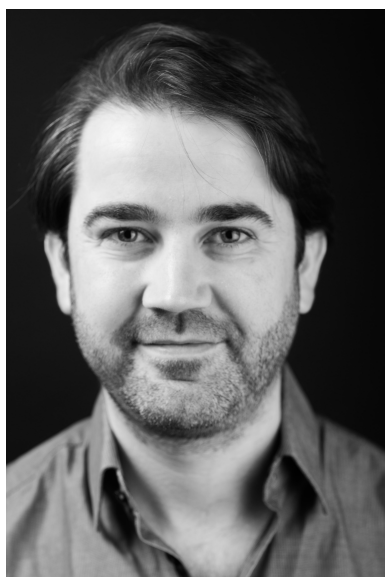

**Rodin T. A. Aarssen** is a PhD candidate in the Software Analysis and Transformation group at Centrum Wiskunde & Informatica (CWI) and Eindhoven University of Technology. His research focuses on meta programming techniques to perform analyses and transformations on existing software. He can be reached at Rodin.Aarssen@cwi.nl.

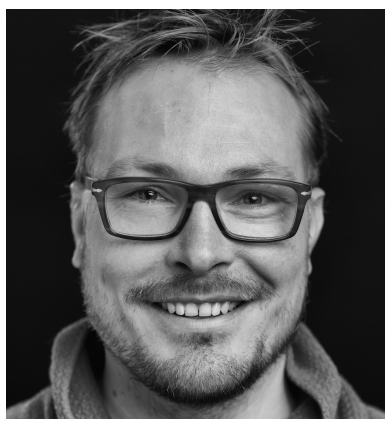

**Jurgen J. Vinju** is full professor of Automated Software Analysis at Eindhoven University of Technology, research group leader at Centrum Wiskunde & Informatica (CWI), and senior language engineer and co-founder of SWAT.engineering. He studies the design and evaluation of (applications of) meta programming systems to get the complexity of source code maintenance under control. Examples are metrics and analyses for quality control or debugging, and model driven engineering for code generation. For more information, see http://www.cwi.nl/~jurgenv. He can be reached at Jurgen.Vinju@cwi.nl.

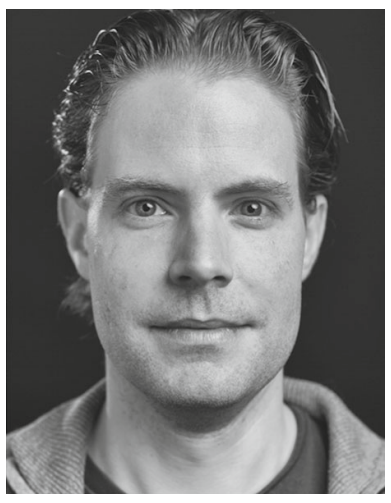

**Tijs van der Storm** is senior researcher in the Software Analysis and Transformation group at Centrum Wiskunde & Informatica (CWI), and full professor in Software Engineering at the University of Groningen. His research focuses on improving programmer experience through new and better software languages and developing the tools and techniques to engineer them in a modular and interactive fashion. For more information, see http://www.cwi.nl/~storm. He can be reached at T.van.der.Storm@cwi.nl.